# Supercontinuum generation and carrier envelope offset frequency measurement in a tapered single-mode fiber

Long Zhang,[1] Hainian Han,[1, a] Yanying Zhao,[2] Lei Hou,[1] Zijiao Yu,[1] Zhiyi Wei[1, b]

[1] Beijing National Laboratory for Condensed Matter Physics, Institute of Physics,
Chinese Academy of Sciences, 100190, Beijing, China
[2] State Key Laboratory of Nuclear Physics and Technology,
Peking University, 100871, Beijing, China

[a] Corresponding author email: hnhan@iphy.ac.cn
Tel: +86-10-82648178; Fax: +86-10-82648178
[b] Email:zywei@iphy.ac.cn

**Abstract** We report supercontinuum generation by launching femtosecond Yb fiber laser pulses into a tapered single-mode fiber of 3 μm core diameter. A spectrum of more than one octave, from 550 to 1400 nm, has been obtained with an output power of 1.3 W at a repetition rate of 250 MHz, corresponding to a coupling efficiency of up to 60%. By using a typical $f$-$2f$ interferometer, the carrier envelope offset frequency was measured and found to have a signal-to-noise ratio of nearly 30 dB.

## 1 Introduction

Spectral broadening based on nonlinear effects has received widespread attention due to its rich spectral content, excellent coherence and high spatial brightness. It has a broad range of applications in areas such as optical fiber communication [1], optical coherence tomography [2] and frequency metrology [3]. Of particular importance is its use in the measurement of the carrier envelope offset frequency ($f_{ceo}$) of femtosecond optical frequency combs, wherein a spectrum extending over one optical octave is necessary when employing the $f$-$2f$ interferometer technique.

It is usual to use a photonic crystal fiber (PCF) [4-9] to broaden the laser spectrum. Because of its special structure it is easy to confine light in the small PCF core area and customize the dispersion characteristics. Broadband spectra of more than one octave have been generated with ultrashort laser pulses in PCF. An alternative scheme for spectrum broadening is fiber taper technology, which was first applied in conventional single-mode fiber, and has since remained quite popular [10-14]. Compared with PCF, the tapered single-mode fiber has a natural double-sided funnel structure which enables light to be focused much more easily and thus generate a supercontinuum (SC) with higher efficiency and stability. Tapered single-mode fibers may therefore have great potential in developing state-of-the-art frequency combs, which demand stable, high power SC generation for reliable performance. On the other hand, the single-mode tapered fiber technique is quite simple, inexpensive, and flexible for various applications. In 2000, Birks *et al.* reported spectrum broadening in a tapered single-mode fiber with a 2 μm diameter in the central portion of a standard telecommunications fiber. As a result, an SC extending across a two-octave spectrum was achieved by using femtosecond laser pulses generated by Ti:sapphire [11] and Cr:forsterite [12] oscillators. In addition, the tapered single-mode fiber is able to support high power transmission up to several watts. For example, a multi-watt SC source was set up in cascaded tapered single-mode fibers of different core diameters [13]. An average output power of more than 5 W was obtained. When the core diameter is reduced to less than 1.5 μm, the zero dispersion wavelength (ZDW) matches well with visible light, which is beneficial for the generation of white-light SC [14]. Although many important applications, such as mode propagation, four-wave mixing and optical sensing have been realized by using tapered single-mode fibers [15-18], to our knowledge no measurements of the $f_{ceo}$ in an optical frequency comb have been reported before.

In this paper, we describe the generation of a SC and the measurement of $f_{ceo}$ by injecting high repetition rate femtosecond pulses into a tapered single-mode fiber. The laser pulses, generated by an ytterbium fiber (Yb-fiber) oscillator, were amplified and compressed to nearly 115 fs. An output power up to 2.2 W was achieved, corresponding to a single pulse energy of 8.8 nJ at a repetition rate of 250 MHz. Then the amplified pulses were coupled into a home-made tapered single-mode fiber with a core diameter of 3 μm and taper length of 9 cm. Because of the strong nonlinear effects in the tapered fiber, a spectrum covering over an octave, from 550 to 1400 nm, was obtained. The SC power was more than 1.3 W with a coupling efficiency of up to 60%. Based on this SC spectrum, the $f_{ceo}$ was measured with a typical $f$-$2f$ interferometer, which indicates that the tapered single-mode fiber could be a suitable choice for constructing optical frequency combs.

## 2   Experiment

### 2.1   *Properties of the tapered single-mode fiber*

The tapered fiber we used was made of a conventional single-mode fiber (SMF-28, Corning). After synchronous heating and pulling in a taper-drawing rig, a tapered section in the center of untreated single-mode fiber was fabricated. In our experiment, the fiber cladding was tapered from 125 to 3 μm over a length of about 9 cm. The transition length on either side of the tapered section was nearly 3.5 cm, so the total length of the tapered single-mode fiber was 16 cm. When light travels in the tapered section, it propagates within the whole cladding, rather than being confined to the core only. Because of the large index difference between the cladding and air, the light is guided quite well in the tapered fiber. In order to avoid touching other objects, the fiber was hung and enclosed in a plastic housing, to ensure low-loss transmission and easy handling.

According to previous analyses [19-21], the nonlinear coefficient largely depends on the fiber parameters, such as the core diameter, core-cladding index difference, and the effective core area of the transmission mode. When we only consider the fundamental mode transmission, the nonlinear coefficient is calculated to be about 53 $W^{-1}km^{-1}$ at a wavelength of 1030 nm.

The dispersion characteristics of the fiber were also estimated. The change of ZDW with core diameter is given in Fig. 1 (a), while the group velocity dispersion (GVD) and third order dispersion in the 3 μm core diameter tapered fiber are shown in Fig. 2 (b). The GVD was about - 46 $ps^2$/km at 1030 nm. In this case, the third order dispersion is very small and has little impact on the nonlinear interaction between the femtosecond pulses and fiber, so it may be neglected.

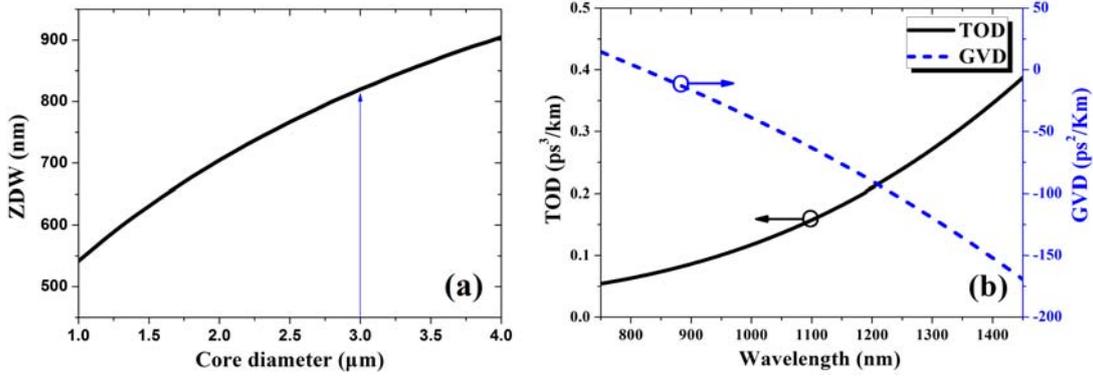

Fig. 1 (a): Zero dispersion wave as a function of core diameter in the tapered single-mode fiber. (b): Calculated group velocity dispersion (GVD, blue, dashed curve) and third order dispersion (TOD, black, solid curve) versus wavelength for a core diameter of 3μm.

Although it is obvious that a smaller core can enhance the nonlinearity and shift the ZDW to shorter wavelengths, it also restricts the spectrum range because of the large difference between the incident and zero dispersion wavelengths. In addition, when the core diameter is too small (< 2 μm), the fiber cannot sustain high power pulses, so we chose a 3 μm tapered fiber to generate SC in our experiment.

## 2.2 Experimental setup

The experimental setup is shown in Fig. 2. It consists of a 250 MHz femtosecond Yb-fiber oscillator (MenloSystems GmbH) and operates in the nonlinear polarization rotation mode-locked configuration. The spectral width is about 50 nm at the center wavelength of 1030 nm, as shown by the blue dashed curve in Fig. 3 (a). The laser output, emitted from an intra-cavity polarization beam splitter, is coupled into a fiber splitter and divided into three routes. One beam has nearly 30 mW average power and is led into a double-stage fiber amplifier. Optical isolators are inserted between the oscillator and amplifiers to avoid unnecessary damage caused by detrimental reflected light. After amplification, the pulses are compressed by using a pair of transmission gratings. Due to the influence of gain narrowing during amplification, the intensity around the emission peak will be amplified effectively, so the spectral width after compressing is narrowed to about 15 nm, as shown by the black solid curve in Fig. 3 (a). The intensity autocorrelation curve (Fig. 3 (b)) shows that the pulse duration is about 115 fs for an amplified power of 2.2 W.

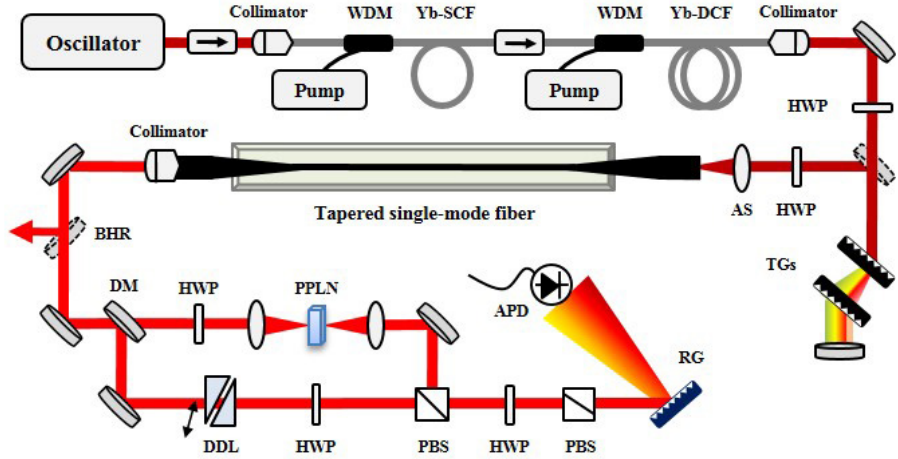

Fig. 2 Experimental setup of SC and $f_{ceo}$ generation with a tapered single-mode fiber. AS: aspherical lens, Yb:SCF/DCF: Yb doped single-clad / double-clad fiber, HWP: half-wave plate, TGs: transmission gratings, BHR: broadband half reflective mirror (coating: 900 ~ 1200 nm), DM: dichroic mirror, DDL: dispersion delay line, PBS: polarization beam splitter, RG: reflection grating, APD: avalanche photodetector.

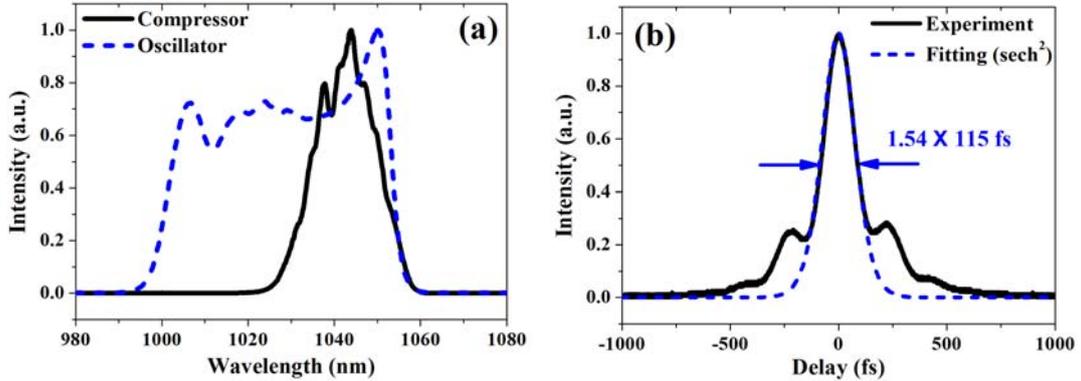

Fig. 3 (a) Normalized spectral intensity of the oscillator (blue dashed curve) and compressor (black solid curve). (b) Intensity autocorrelation curve after compressing (black solid curve) and $sech^2$-fitting result (blue dashed curve).

## 3 Results and discussion

### 3.1 Supercontinuum generation

In order to obtain an octave-spanning spectrum, we used the 3 μm core diameter, 9 cm long tapered single-mode fiber described above. The enclosed fiber was mounted on a three-dimension precision stage. The compressed pulses were focused into the fiber with an aspherical lens of 6.24 mm focal length, and the output was collimated with an aspherical lens attached to a FC/APC patch cable. The output was measured with an optical power meter and spectrum analyzer (AQ 6315A, ANDO). The SC output as a function of input power is shown in Fig. 4 (a), where we see that the maximum transmitted power is 1.3 W for an input power of 2.2 W. The coupling efficiency is about 60%, and stable operation can be maintained for a long time by reason of the large core diameter at the fiber end surfaces which do not have a honeycomb structure.

Figure 4 (b) shows the change in the SC spectrum for different input pulse energies, the values of which are indicated on the right vertical axis. It is evident that the spectral bandwidth broadens as the coupled energy increases. Nonlinear effects, such as soliton fission, Raman scattering and dispersion effects, play important roles during the evolution process [6]. When the pulse energy is less than 0.8 nJ, soliton fission plays a dominant role in the abnormal dispersion regime. As s result, symmetric broadening occurs and, accordingly, high order solitons are generated around 1030 nm. As the pulse energy increases, Raman perturbation will affect soliton production which will shift towards longer wavelengths, extending the spectrum from 1100 to 1400 nm [22]. On the other hand, visible components will gradually appear under the impact of dispersive wave radiation [23, 24]. The discrete peak around 600 nm is first observed at a coupled energy of 1.6 nJ. When higher pulse energies are injected, more spectral components are induced between 550 to 900 nm. The peak around 480 nm might be the second harmonic generated from the residual pump laser beam. Finally, at the maximum pulse energy of 5.2 nJ, the SC span exceeds an octave, extending from 550 to 1400 nm. However, the shape of the SC spectrum is rather ragged across its range. The reason for this is probably that the ZDW of the tapered fiber we used is slightly offset from the incident wavelength, which results in more pulse energy being distributed to the near infrared region. In spite of this, the broadened spectrum can span more than one octave range and completely satisfy the needs for the high and low frequency components of $f_{ceo}$ detection in an $f$-$2f$ interferometer.

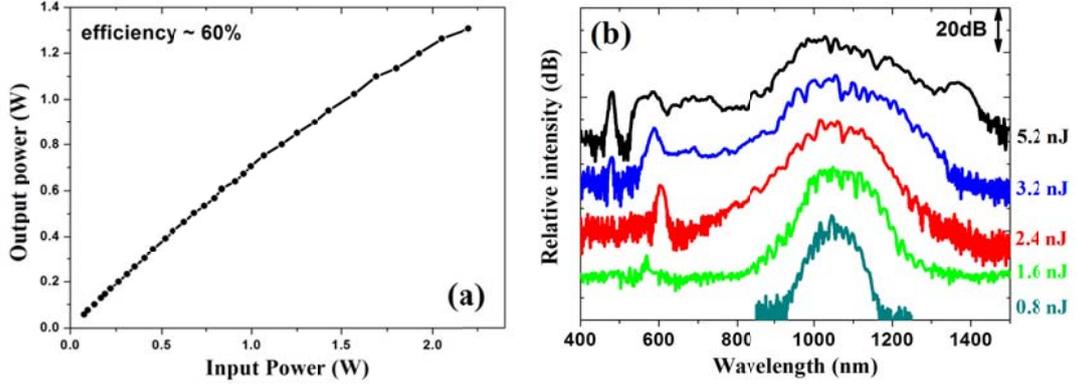

Fig. 4 (a) Supercontinuum output as a function of input power; the coupling efficiency is about 60 %. (b) Supercontinuum spectrum for different coupled pulse energies from 0.8 to 5.2 nJ.

### 3.2 *Measurement of carrier envelope offset frequency*

Here we choose the SC spectrum corresponding to a pulse energy of 3.2 nJ for $f_{ceo}$ detection. One part of the SC beam is exported with a 50% broadband reflective mirror coating from 900 to 1200 nm for other applications. The remaining is delivered into a typical $f$-$2f$ interferometer. The long-wavelength component around 1156 nm is frequency-doubled, and beats with the short-wavelength around 578 nm in the SC. A dispersion delay line consisting of two rectangular prisms is placed in the short-wavelength path. The $f_{ceo}$ of the Yb-fiber system is determined by the heterodyne beat signal, which is detected by an avalanche photoelectric detector in the visible overlap region after a diffraction grating. Thanks to the good stability of the fiber ring cavity configuration, the $f_{ceo}$ center frequency is very stable with a resolution band width of 100 kHz over a long period of time. In this regard, it is much better than a Ti:sapphire-based comb whose $f_{ceo}$ usually drifts several megahertz in a minute. However, the linewidth is not as good. As shown in Fig. 5, the free running 3 dB width of the $f_{ceo}$ is about 500 kHz in our experiment, which is a little larger than that in previous results. The main reason for this comes from the dispersion characteristics of the Yb-fiber source. It is impossible to change the intra-cavity net dispersion in a commercial oscillator, which is a crucial factor in determining the $f_{ceo}$ noise. The narrowest linewidth always occurs in the case of near zero-dispersion [25]. In addition, the high power fiber amplifier may also worsen the quality of $f_{ceo}$.

Improvements in optimizing the dispersion and amplifier configuration should be helpful in reducing the noise and increasing the $f_{ceo}$ signal-to-noise ratio (SNR).

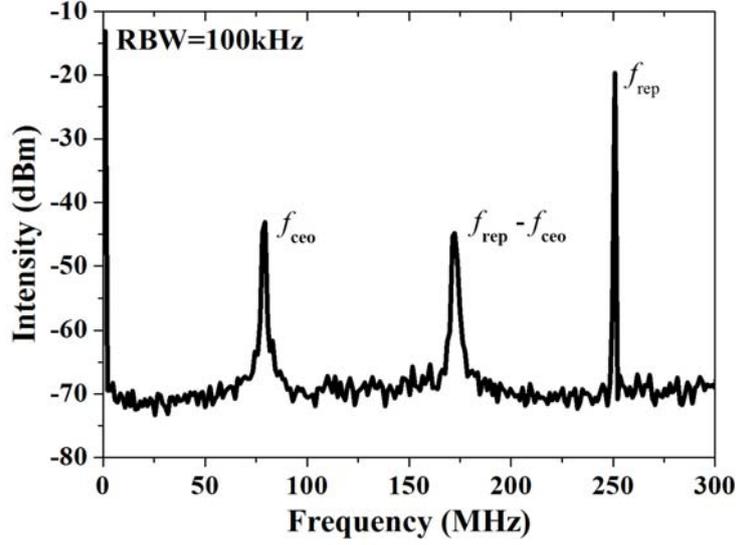

Fig. 5 The $f_{ceo}$ measurement with a typical $f$-$2f$ interferometer, $f_{rep}$ is the repetition rate.

## 4  Conclusion

In conclusion, we have reported the generation of a broadband supercontinuum spanning more than one octave in a tapered single-mode fiber with a core diameter of 3 μm. The spectrum ranges from 550 to 1400 nm, and an SC power of over 1.3 W is obtained with a coupling efficiency of nearly 60%. Based on such a broadband SC, the $f_{ceo}$ signal, which has an SNR of nearly 30 dB and line width of 500 kHz, has been characterized experimentally. Our results indicate that this tapered single-mode fiber could be applied in $f_{ceo}$ measurements and thus used to construct optical frequency combs. Inexpensive fabrication and the possibility to customize parameters are promising advantages for our optical frequency comb system, allowing more practical and stable operation in real application fields.

**Acknowledgments**

We thank Prof. Wei Ding for helpful discussions. This work is partially supported by the National Basic Research Program of China (973 Program Grant No. 2012CB821304), and the National Natural Science Foundation of China (Grant Nos.